# Parametric study of helicon wave current drive in CFETR

Xianshu Wu[1], Jingchun Li[1*], Jiale Chen[2], Guosheng Xu[2], Jiaqi Dong[3], Zhanhui Wang[3], Aiping Sun[3] and Wulv zhong[3]

[1]Department of Earth and Space Sciences, Southern University of Science and Technology, Shenzhen 518055, People's Republic of China.
[2]Institute of Plasma Physics, Chinese Academy of Sciences, Hefei, Anhui, China
[3]Southwestern Institute of Physics, Chengdu 610041, People's Republic of China.
*Corresponding author: Jingchun Li (jingchunli@pku.edu.cn)

**Abstract.** This paper evaluates the feasibility of helicon current drive (HCD) in a hybrid scenario for the China Fusion Engineering Test Reactor (CFETR). Utilizing the GENRAY/CQL3D package, a large number of simulations (over 5 000) were conducted, with parametric scans in the antenna's poloidal position, launched parallel refractive index ($n_{\parallel}$), and wave frequency. The analysis reveals that helicon has excellent accessibility under reactor-level conditions, and smaller $n_{\parallel}$ and higher wave frequency result in enhanced wave absorption. The simulations demonstrate an optimal launched $n_{\parallel}$ of approximately 1.6 for the CFETR hybrid scenario, with helicon achieving a maximum drive efficiency of $2.8 \times 10^{19}$ A·W$^{-1}$·m$^{-2}$. The best launch position is found to be within a poloidal angle range of 25 degrees to 65 degrees. Additionally, it is preferable to have a narrow $n_{\parallel}$ spectrum for wave absorption when operating below the threshold value of $\Delta n_{\parallel}$ (~0.6), beyond which the effect of $\Delta n_{\parallel}$ on wave absorption is negligible. This study provides valuable insights into the potential application of HCD in CFETR.

## 1. Introduction

Fast waves in magnetized plasmas can be broadly classified into two types. The first type, commonly referred to as "fast waves" in tokamaks, has a frequency in the ion cyclotron frequency range. The second type is known as the "helicon" wave and has a much higher frequency, close to the lower hybrid frequency. This second type of fast wave is also referred to as a "whistle wave" or "fast wave in the lower hybrid frequency range". The propagation of the first type of fast wave is predominantly influenced by ion absorption in the ion cyclotron resonance layer. In contrast, the second type, the "helicon" wave, mainly experiences absorption by electrons through the process of Landau damping [1].

Helicon current drive, also known as fast wave current drive in the lower hybrid range of frequencies, is regarded as a promising tool for driving off-axis currents in reactor-grade plasmas. In future steady-state fusion devices, it could provide a solution to the drawbacks associated with electron cyclotron waves and lower hybrid current drive. The helicon wave system was first installed and tested on the JFT-2M tokamak using a 12-element combline antenna [2]. The traveling wave antennas (TWA) has valuable features such as load resiliency, a narrow $n_\parallel$ spectrum, and a simple radio frequency (RF) circuit that doesn't require external matching systems. The combline TWA antenna exhibited the expected characteristics during the JFT-2M experiments. The input impedance matched well with the transmission line impedance under various plasma loads, including H-mode, L-mode, ohmic, and vacuum (for pre-ionization) conditions. The antenna operated reliably within the available power range of 400 kW and showed no signs of power saturation. The measured wave electric fields at the plasma center were consistent with the full-wave simulations, indicating successful excitation of the fast wave[3].

Currently, mid-sized to large tokamaks such as KSTAR and DIII-D are equipped with helicon systems. In KSTAR, a low power (mW) mock-up combline-type TWA was tested during discharges, and various plasma conditions such as L/H modes were studied to measure changes in impedance matching, coupling, and dominant $n_\parallel$[4]. To distinguish between slow and fast wave coupling, the magnetic field tilt angle was scanned. The experimental and analytical results showed that the fast wave dominated in relatively high coupling, and the use of the TWA antenna was found to provide load flexibility. To achieve practical off-axis current drive applications, KSTAR is preparing to install a 1 MW helical wave current drive system[5]. A 30-module traveling wave antenna has been installed and optimized in-vessel at DIII-D in early

2020, achieving good performance in the 10 MHz band around 476 MHz, with approximately 2% reflected power and approximately 1.5% dissipated power per module. The main elements of the design, construction, installation, and commissioning of the 1.2 MW helical wave system have been successfully completed. At present, this RF system has low reflectivity, low losses, and good antenna directionality. The physical basis for helical wave current drive may soon be validated through this system. [6, 7]

Theoretical studies have shown that a helicon current drive (HCD) is highly promising for off-axis current drive in DIII-D, ITER, FNST, and DEMO-grade plasmas [8–12], as well as some spherical tokamaks such as EXL-50 [13]. The primary methods used include linear [14], quasi-linear [8], and full-wave approaches [12]. The full-wave and linear calculations have been well-verified with respect to wave trajectories. However, the extent to which quasi-linear effects impact the results at current device scale remains unclear [8, 15]. It is certain that non-linear effects due to parameter decay cannot be ignored when the wave input power is above 1MW from the current DIII-D experiments. Additionally, recent research has focused on the impact of density turbulence on wave absorption [12], as well as the effects of the scrape-off layer (SOL) region [14, 16]. Therefore, although there has been numerous theoretical and simulation studies on helicon heating and drive, many aspects of helicon physics remain unclear and require further investigation.

This paper presents a scoping study on the potential use of HCD for the China Fusion Engineering Test Reactor (CFETR). One of the potential operating scenarios for CFETR, namely the "hybrid" scenario in which some of the plasma current is sustained by the ohmic transformer, is chosen [17]. The equilibrium reconstruction code EFIT [18] provides the necessary GEQDSK file, while GENRAY [19] and CQL3D [20] solve the ray tracing and Fokker-Planck equations, respectively. The CQL3D calculations include quasi-linear effects. In order to perform a parametric study, over 5000 cases have been carried out. For our studies, a fixed operating frequency of 1.4 GHz is assumed, unless otherwise stated.

The structure of this paper is as follows. In Sec. 2, we briefly describe the helicon physics in toroidal plasma. A broad parametric scan of antenna poloidal location and launched $n_\parallel$ for CFETR are presented in Sec. 3. In Sec. 4, we summarize the results.

## 2. Wave accessibility and absorption

To study the characteristics of the propagation of the slow and fast branches, we are considering the linear (small-amplitude) propagation of waves in steady-state conditions, where the parameters of the medium vary only in one direction, which we refer to as the *x*-direction in the Cartesian coordinate system. This direction corresponds to the minor radius direction in circular magnetic confinement devices (such as tokamaks, stellarators, or RFPs). Assuming that the variation of parameters in the *x*-direction is slow enough that the geometrical optics approximation is valid, i.e., $\partial_x = ik_x$, we can obtain the dispersion relation for the Low-Hybrid range of Frequency:

$$(n_x^2 - n_F^2)(n_x^2 - n_S^2) + n_x^2 \left[\frac{n_z D}{n_z^2 - S}\right]\left[\frac{n_z D}{S}\right] = 0, \qquad (1)$$

that means there will be two branches wave exist in LHRF while the secondary term above is not significant. According to ref.[16], the capability of fast wave branch to access the higher density will not be affected by low hybrid resonance if $\omega < \Omega_{LHR}$, while the slow branch $n_\perp^2$ will tend to infinity at resonant density. This is not considering that there is a particular value of $n_\parallel$ will lead a sufficiently large coupling term in Eq. (1), however which will make the two branches indistinguishable. Further analysis shows that higher density will be "inaccessible" for fast wave from the lower density side. This critical value of $n_\parallel$ can be arrived by the general cold plasma dispersion relation:

$$S n_x^4 + [(P + S)(n_z^2 - S) + D^2]n_x^2 + P[(n_z^2 - S)^2 - D^2] = 0, \qquad (2)$$

Which is obtained by multiplying out the local dispersion relation (1). Here we want the dispersion equation give us two different roots that discriminant of this biquadratic should be positive, that is $[(P + S)(n_z^2 - S) + D^2] - 4SP[(n_z^2 - S)^2 - D^2] > 0$. With the densities where the waves are propagating, in LHRF, we have $|P| \approx \left|1 - \frac{\omega_{pe}^2}{\omega}\right| \gg 1$ and $|S|$ is of order 1, so we can replace $P + S$ and $(n_z^2 - S)^2 - D^2$ by $P$ and $D^2$ respectively, and the inequality becomes

$$[(n_z^2 - S)P + D^2]^2 + 4PSD^2 > 0, \qquad (3)$$

then, we have

$$n_z^2\big|_c > \frac{2\sqrt{S}D}{\sqrt{|P|}} + \frac{D^2}{P} + S = \left(\frac{D^2}{\sqrt{|P|}} + \sqrt{S}\right)^2, \qquad (4)$$

This only can offer us an estimation that the accessibility of waves with a fixed $n_z$, however we need to consider the behavior of $n_z^2\big|_c$ from the edge of the plasma, where the density is zero, up to the maximum density. A more accurate method to study the accessibility of waves can be carried out through solving the dispersion relation directly, which is shown in Fig. 1. This figure shows the density range of slow and fast wave propagation effects resulting from solving the cold plasma dispersion relation at fixed parameters. The parameters are $B_0 = 7.2\text{T}, n_{e0} = 1.2 \times 10^{20}$ m$^{-3}, f = 1.4\text{GHz}$, and the ion species is deuterium. To enhance the presentation effect, we use the inverse hyperbolic *sine* function (sinh$^{-1}$) to the quantity $n_\perp^2$ for the ordinate. As can be seen, the slow wave branch can not access the higher densities because of the lower hybrid resonance, while the fast wave branch is unaffected by it. At the parameters that $n_\parallel(n_z) = 2$ or 3, there is no indistinguishable part to make the fast branch inaccessible for the higher densities. In addition, the cut-off density for fast wave is about the order of $10^{19}$m$^{-3}$, which is satisfied with our expectation. Thus, as demonstrated by the above analysis, the helicon branch (or fast wave branch) exhibits excellent accessibility to the mid-radius region from the low-field side.

The absorption of high harmonic fast waves can be calculated from the imaginary part of the wave number

$$k_{\perp i} = \frac{\sqrt{\pi}}{4} k_\perp \beta_e \xi_e e^{-\xi_e^2} G \qquad (5)$$

where

$$G = \frac{n_\parallel^2 - S}{A_2} + \frac{\left[1 + \frac{\omega_{pe}^2 Re(Y)}{\omega_{ce}^2} \frac{1}{|Y|^2} \frac{1}{n_\parallel^2 - S}\right]\left(\frac{mc^2}{kT_e}\right)^2 \frac{\omega_{ce}^2 D^2}{\omega^2 |\epsilon_{33}|^2}}{A_1 A_2} \quad (6)$$

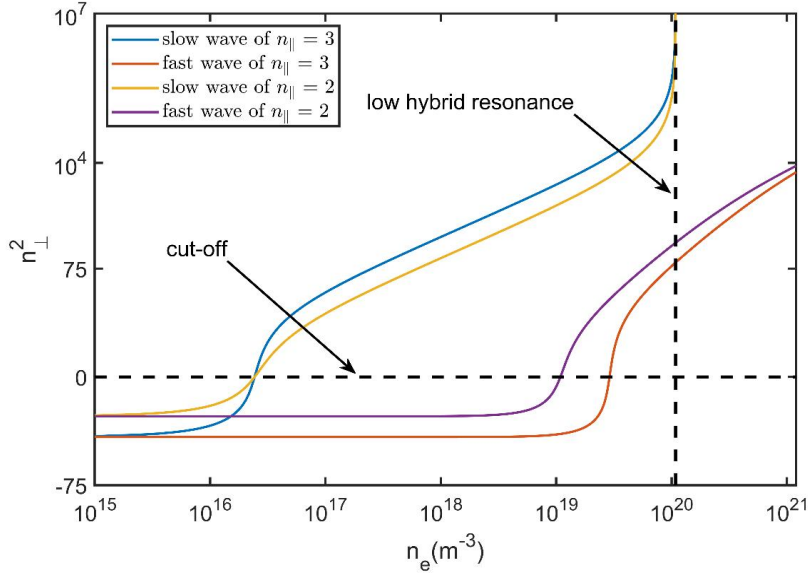

Figure 1 Density ranges of slow and fast wave propagation and the effect of the lower hybrid resonance on the slow branch at different $n_\parallel$, with $B_0 = 7.2\text{T}$, $n_{e0} = 1.2 \times 10^{20} \text{m}^{-3}$, $f = 1.4\text{GHz}$, deuterium.

$$k_\perp = \frac{\omega}{c}\sqrt{\frac{\left(S - n_\parallel^2\right)^2 - D^2}{S - n_\parallel^2 + D^2 \frac{Re(\epsilon_{33})}{|\epsilon_{33}|^2}\frac{n_\parallel^2}{\left(S - n_\parallel^2\right)}}} \quad (7)$$

Here $\omega_{cs} = q_s B/m_s$ is the cyclotron frequency of the particle, $\omega_{ps} = \left(\frac{n_s q_s^2}{m_s \epsilon_0}\right)^{\frac{1}{2}}$ is the plasma frequency, $n_\parallel$ is the parallel refractive index, and $B$ is the magnetic field, $Y = 2\xi_e^2[1 + \xi_e Z(\xi_e)]$. The expressions for the coefficients A1 and A2 are as follows,

$$A_1 = n_\parallel^2 - S + \frac{\omega_{pe}^2}{\omega_{ce}^2} \frac{Re(Y)}{|Y|^2} \frac{n_\parallel^2}{n_\parallel^2 - S}$$

$$A_2 = n_\parallel^2 - S + \frac{\omega_{pe}^2}{\omega_{ce}^2} \frac{Re(Y)}{|Y|^2} \frac{n_\parallel^2 + S}{n_\parallel^2 - S}$$

$\epsilon_{33} = Y\frac{\omega_{pe}^2}{\omega^2}$ is an element of the dielectric matrix. $R, L, S$ is the Stix wave notation

and can be find in ref.[21]. $Z$ is the dispersion function. When considering only real numbers, its expression is as follows

$$Z(\xi_e) = i\pi^{1/2} \exp(-\xi_e^2) - 2\xi_e N(\xi_e)$$

$$N(\xi_e) = \left[\exp(-\xi_e^2)/\xi_e\right] \int_0^{\xi_e} \exp(t^2) \, dt$$

Fig. 2 shows the contour lines of $0.2k_{\perp i}a$, where $k_{\perp i}$ is from equation (5) and $a$ is the minor radius, for parameters characteristic of a hybrid scenario in CFETR that will be described later. The parameters used in the calculations correspond to the location ($\rho$=0.6) where the helical wave is mostly absorbed in the CFETR configuration: $B = B_0 = 7.2$ T, $n_e = 9.2 \times 10^{19}$ cm$^{-3}$, $kT_e = 12$ keV. In Fig.2(a) and (c), $f = 1.4$ GHz, while in Fig.2(b) and (d), $\beta_e = 0.02$. The parameter $0.2k_{\perp i}a$ characterizes the wave absorption, so it can be used to represent the absorption of the wave within the small radius. When $0.2k_{\perp i}a$ is greater than 1 (indicated by bold lines), the wave is absorbed within a small range of the small radius. Within the parameter range shown in the figure, Fig1(a)(c) indicates that absorption becomes stronger as $\beta_e$ increases. Fig2(b)(d) shows that for low wave frequencies, absorption becomes stronger as the frequency increases. When the wave frequency is approximately greater than 1.1 GHz, the absorption rate does not change significantly with frequency. In addition, absorption increases as $\xi_e$ decreases. From (a-c) or (b-d), it can be observed that when $n_\parallel$ increases from 2 to 3, the contour lines "shrink" towards the lower right corner, indicating a decrease in absorption.

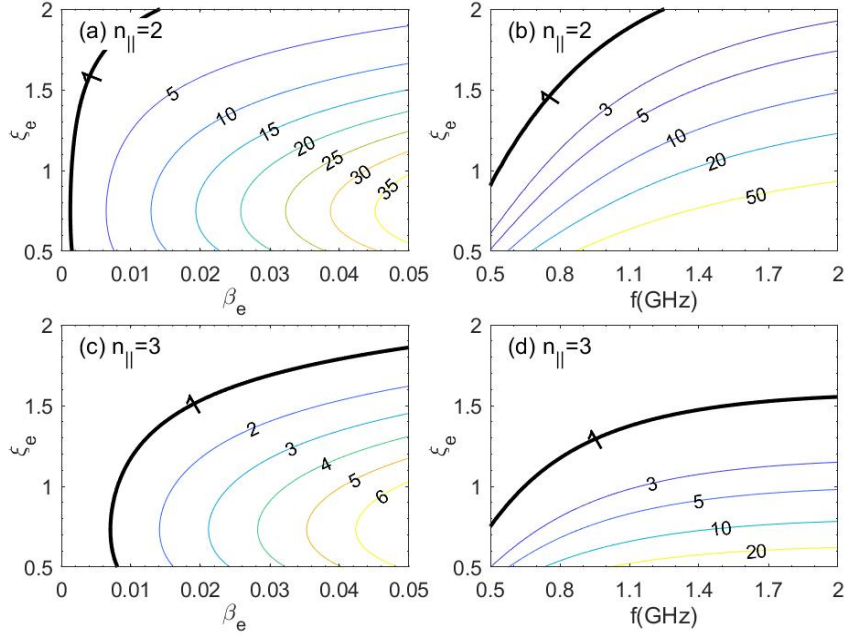

Fig 2. (a) Contour of $0.2K_{\perp i}$ as a function of $\beta_e$ and $\xi_e$ with wave frequency fixed at 1.4GHz, (b) Contours of the same quantity as a function of the applied frequency and $\xi_e$, with $\beta_e$ fixed at 2.0%.

## 3. Helicon wave heating and current drive in CFETR

In this section, we investigate the capability of high harmonic fast wave heating and current drive using the GENRAY and CQL3D packages. GENRAY is a linear ray-tracing code that solves the wave trajectory equation, while CQL3D incorporates the quasi-linear effect and collisionality. In our simulations, we use the CFETR hybrid scenario, which has a plasma temperature and density as shown in Figure 3, with a core temperature of 30KeV and core density of $1.2\times10^{20}m^{-3}$. The input power for all cases is 10MW, and the cold plasma dispersion relationship is adopted in the ray-tracing calculations.

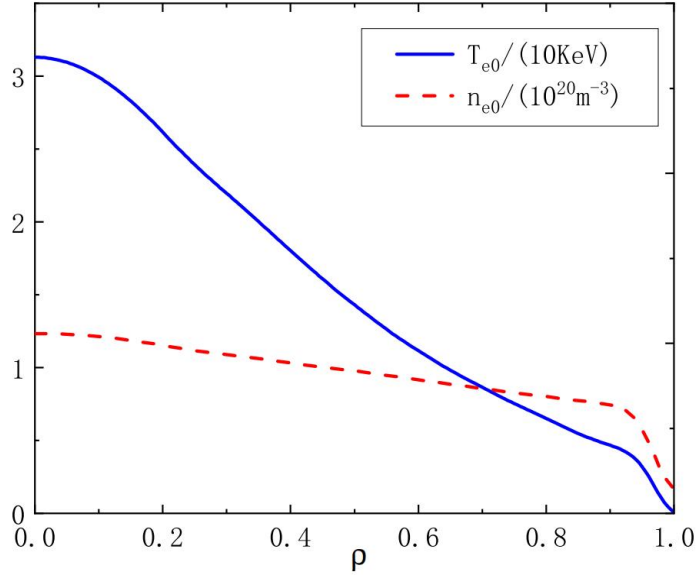

Fig 3. The temperature and density profile for the CFETR hybrid scenario

A typical run of helicon heating and current drive with the GENRAY/CQL3D code is depicted in Figure 4, with a poloidal injection angle of 5°. Figures 4(a-b) represent the wave trajectories in both real and phase space, while Figures 4(c-d) demonstrate the power absorption density and driven current density. We observe that the waves slowly spiral radially inwards, gradually depositing their energy. The peak of the deposition profile is located at $\rho$=0.51. The total current driven by helicons is 360KA, which corresponds to a current drive efficiency of $\eta_{helicon} = n_e IR/P_{helicon} = 2.4\times10^{19}$ A·W$^{-1}$·m$^{-2}$.

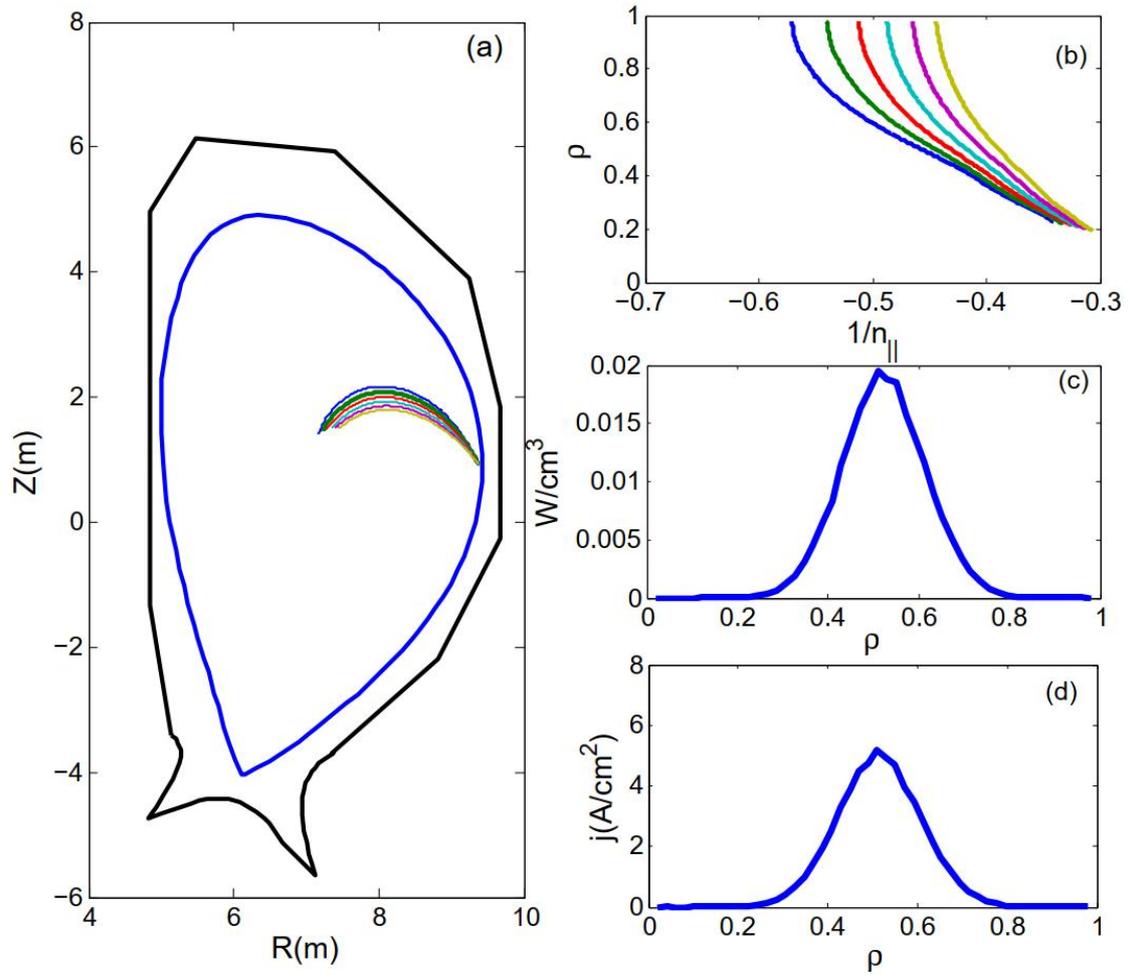

Fig 4. (a) Trajectories of helicon in CFETER configuration, (b) helicon wave trajectories in the phase space, (c) power absorption density, and (d) driven current density with poloidal angle of 5° and $n_{||}$ of -2. The total driving current is 360KA.

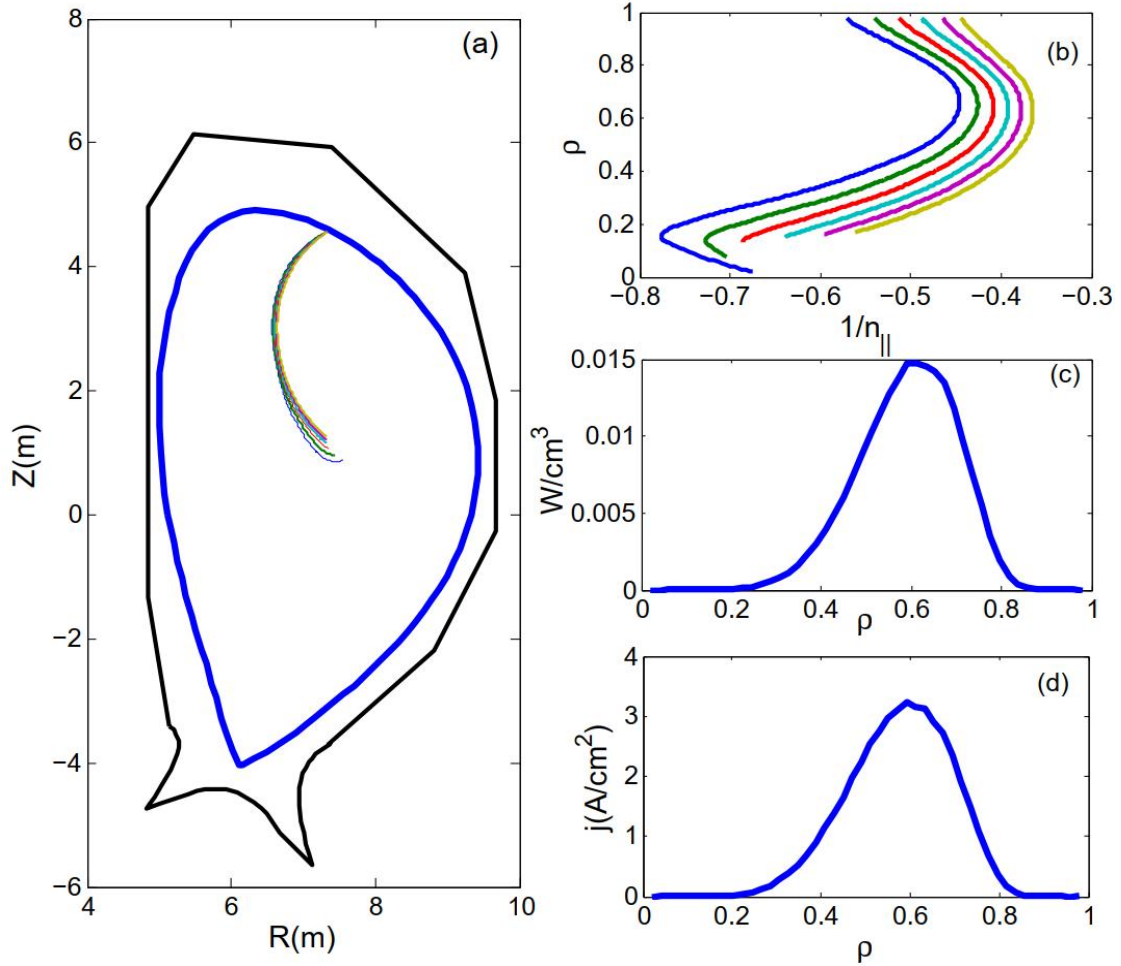

Fig 5. (a) Trajectories of helicon in CFETER configuration, (b) helicon wave trajectories in the phase space, (c) power absorption density, and (d) driven current density with poloidal angle of 95° and $n_{\|}$ of -2. The corresponding total driving current is 326KA.

Figures 5 and 6 summarize the simulation results at the optimal launched parallel refractive index ($n_{\|}=-2$) for emission positions near the plasma top (95°) and the high-field side (185°). It can be observed that the emission on the high-field side yields weaker current drive results compared to the top emission and even lower than that of the midplane emission (as also evident from Figure 7). This is in contrast to the conclusion that emission on the high-field side with lower hybrid wave can achieve higher drive efficiency. For lower hybrid wave, emission on the high-field side should be optimized at lower n|| to better utilize wave accessibility. Lower $n_{\|}$ allows the wave to penetrate deeper into the higher temperature plasma core (as shown in Equation (1)), which is inaccessible from the low-field side due to strong Landau damping of high $n_{\|}$ waves at the edge. However, for the high harmonic fast wave, the strong damping of the wave also depends on $\xi_e$, and considering the electron temperature, it is important to note that Figure 2 shows that local $\xi_e \approx 0.8$ is the condition for the strongest absorption.

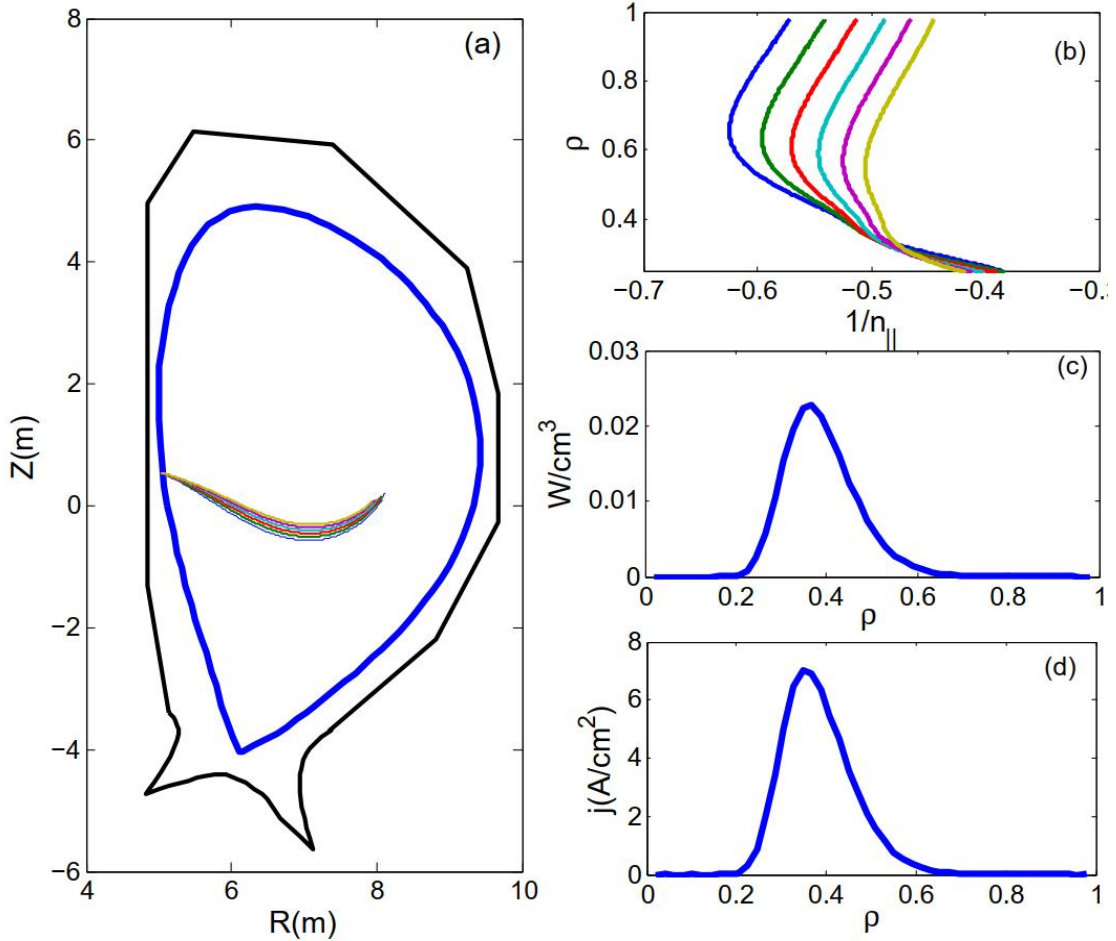

Fig 6. (a) Trajectories of helicon in CFETER configuration,(b)helicon wave trajectories in the phase space, (c) power absorption density, and (d) driven current density with poloidal angle of 185° and $n_{\|}$ of -2. The total driving current is 318KA.

At a fixed frequency of 1.4 GHz, we further scanned two key parameters: the parallel refractive index and the launch angle. Figure 7 presents the scanning results of the peak position of the driven current and the current driven by helicons in the two-dimensional polar coordinate system. The polar angle determines the position where the wave is launched, thus well representing the poloidal injection angle. On one hand, we can clearly see that there exists an optimal parallel refractive index ($n_{\|}$) of about 1.6 for the best wave absorption, which is reasonable. Theoretically, according to equation (5), for wave propagation, absorption, and current drive (CD) efficiency, the key parameter is $\xi_e$, or $n_{||}$, for a target plasma with a specified electron temperature. There exists an optimal $\xi_e$ ($n_{||}$) value due to the competition between two effects: higher $\xi_e$ leads to higher CD efficiency per absorbed power, but on the other hand, power absorption decreases exponentially with $e^{-\xi_e^2}$. Therefore, both too high or too low $n_{||}$ are not favorable for wave absorption. From a physical point of view, absorption weakens as $n_{||}$ parallel to the wave increases, and the wave deposits its energy and momentum in fast electrons near the wave phase velocity $v_{||} = c/n_{||}$. Electrons with higher velocity are less likely to undergo collisions, thus lower $n_{||}$ should lead to higher CD efficiency. However, as we analyzed previously, too low $n_{||}$ will cause bad accessibility of the wave.

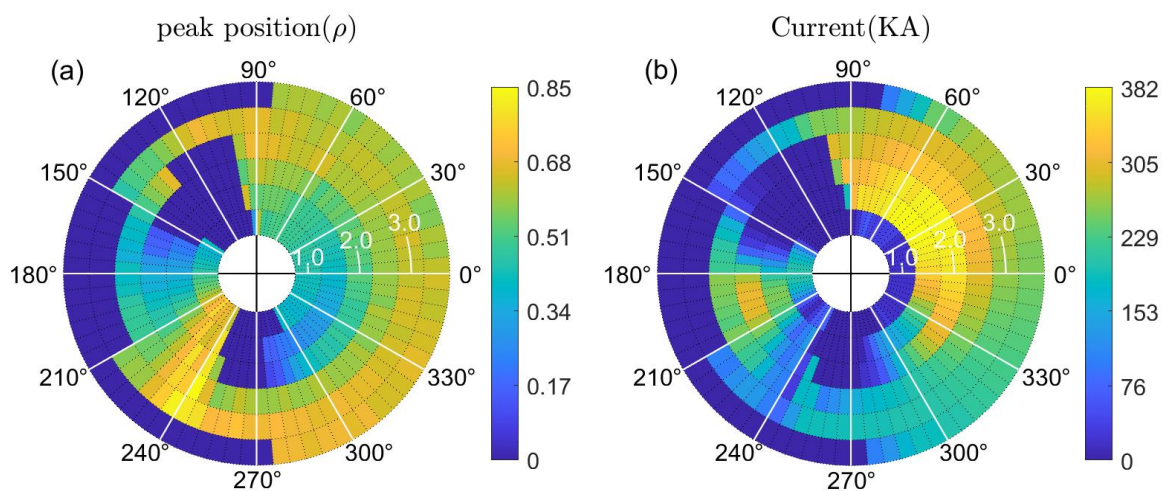

Fig 7. (a)the peak position of driven current and (b) the current driven by helicons in the two-dimensional polar coordinate system. The polar angle indicate the plodial injection angle and the white line represents the value of $n_{||}$.

In addition, we find that the optimal launch position has a poloidal angle range from 25 degrees to 65 degrees, which is very similar to the cases in the DIII-D tokamak where a poloidal angle of 45 degrees for the antenna location is a good choice [8]. Furthermore, when the driven current is at its peak, the current density profile peaks at around $\rho$=0.5-0.6 as shown in Figure 7(a). The key to off-axis current drive is to have plasma conditions that allow the wave to travel toward the plasma center at a not too fast speed, while also having strong enough absorption to sufficiently damp the wave before it gets too close to the plasma axis. From our simulations, we find that too close to the center or the edge of the plasma, the absorption efficiency is greatly reduced.

We conducted further scoping studies on helicon heating and current drive for CFETR. The simulation results of a parametric scan in the launched parallel index of refraction ($n_{||}$), and antenna poloidal position for the CFETR hybrid scenario are shown in Figure 8. From Figure 8(a), we still observe an optimal $n_{||}$(~1.6), at which the driven current can reach 414 kA, corresponding to a current drive efficiency of 2.8 $\times$ $10^{19}$ A·W$^{-1}$·m$^{-2}$. This implies that HCD is highly promising under reactor-relevant conditions, as the highest achievable drive efficiency for lower hybrid waves under the same conditions is only 4.0×$10^{19}$ A·W$^{-1}$·m$^{-2}$ [21]. Another interesting phenomenon is that a small $n_{||}$ results in wave deposition over a wider range of profile peak, from about $\rho$=0.2 to $\rho$=0.7. This is consistent with the fact that a lower $n_{||}$ can allow the wave to penetrate more deeply into the hotter core region of the plasma, resulting in a wider radial range of peak value. Regarding the launch position, upward launch is better than downward launch, and the main deposition range for upward launch is approximately at $\rho$=0.4 to $\rho$=0.65.

The previous calculations were based on a frequency of 1.4 GHz, which is currently feasible with available klystrons. However, future advancements in technology may enable higher operating frequencies. We can perform a simple calculation to evaluate the absorption behavior of the helical wave at higher frequencies. The impact of wave frequency on wave absorption and current drive is presented in Figure 9. The plot displays the power absorption and current drive as a function of wave frequency for a given poloidal emission position and $n_{||}$. It is observed that over a wide frequency range (from 1.4 GHz to 5.5 GHz), wave absorption exhibits a strong dependence on frequency, where higher frequencies result in increased absorption and current drive. This finding is consistent with the analytical results presented in Figure 2. Nevertheless, the frequency has little impact on the deposition location, with only a slight inward shift observed for frequencies above 2.45 GHz. In general, at higher frequencies, the wave trajectory tends to spiral slightly upward and is absorbed relatively deeper inside the plasma, resulting in higher absorption efficiency. In contrast, at lower frequencies, the wave cannot spiral as deeply into the plasma and is absorbed with lower efficiency.

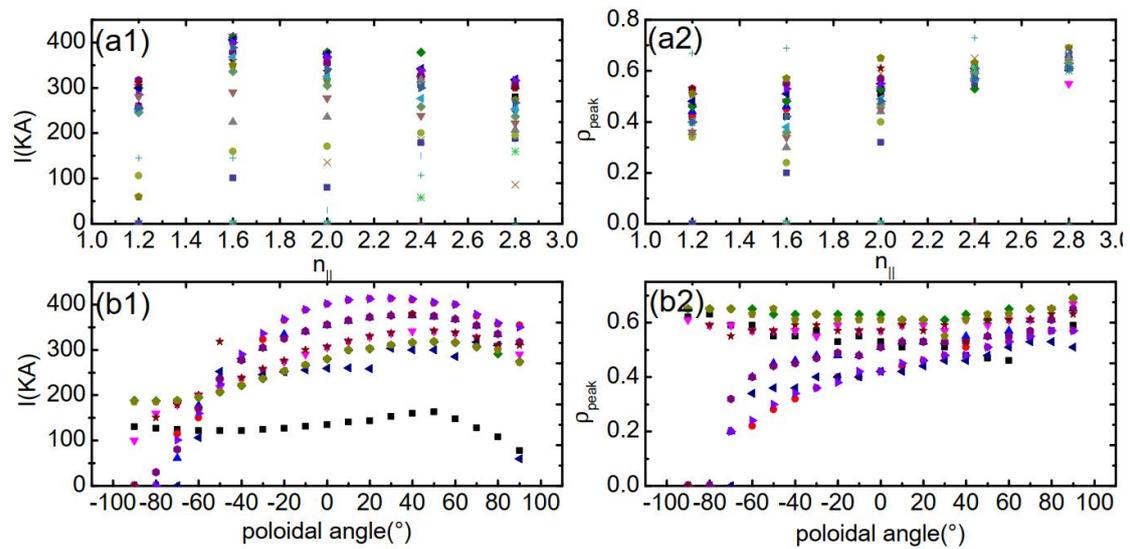

Fig 8. Simulation results of a parametric scan in wave frequency, $f_0$, launched parallel index of refraction, $n_{||}$, and antenna poloidal position for the CFETR hybrid scenario. The (a1-b1, a2-b2) plots represent simulations at $f_0$=1.4GHz by scanning $n_{||}$ and poloidal position.

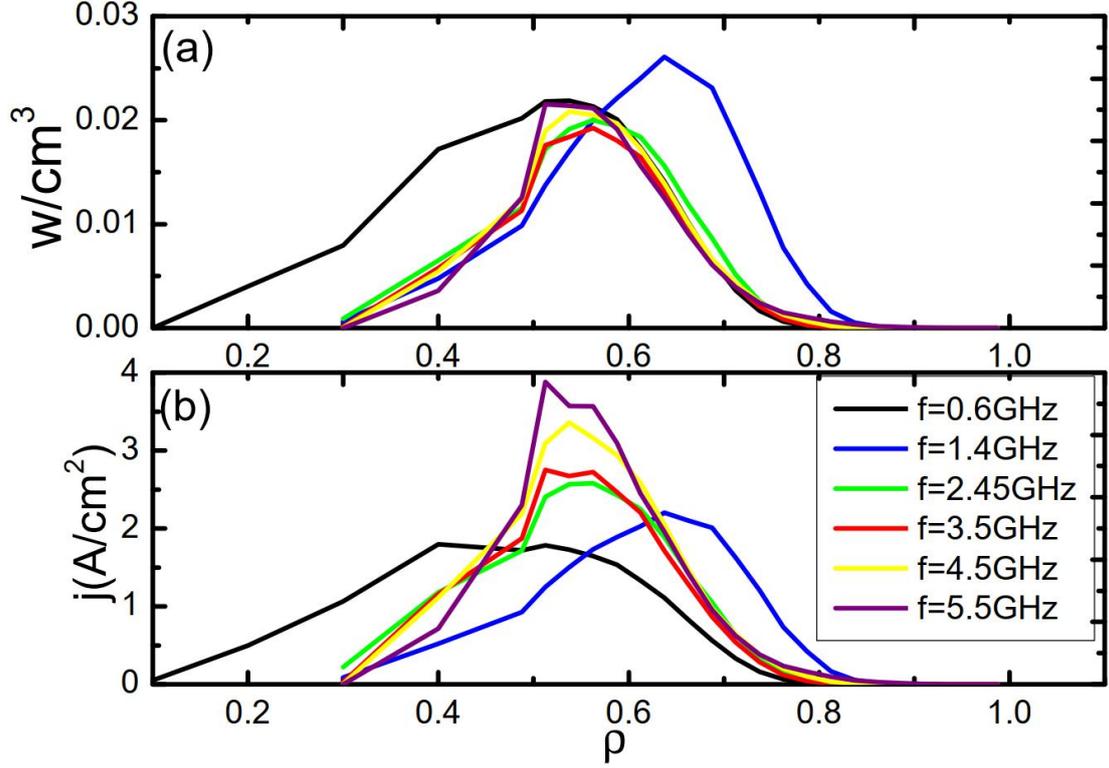

Fig 9. (a)power absorption density, and (d) driven current density with poloidal angle of 90° and $n_{||}$ of -1.6 under different wave frequencies. The driven current is 240.KA, 253KA, 262KA, 253KA, 293KA, and 300KA respectively.

In the GENRAY calculations, we employed an $n_{||}$ spectrum with a full width ($\Delta n_{||}$). The results of a parametric scan in the $n_{||}$ spectrum full width ($\Delta n_{||}$) are depicted in Figure 10, in which we fixed the frequency (1.4 GHz) and $n_{||}$(1.6). From Figure 10, it can be observed that within a certain range ($\Delta n_{||}<0.5$), a more localized $n_{||}$ spectrum can drive higher current. As $\Delta n_{||}$ is further increased, there are no significant changes in power absorption and current profiles. Overall, a more localized $n_{||}$ spectrum is more favorable for wave absorption, and this is relatively easy to achieve for the traveling wave (combline) antenna of the helicon wave system.

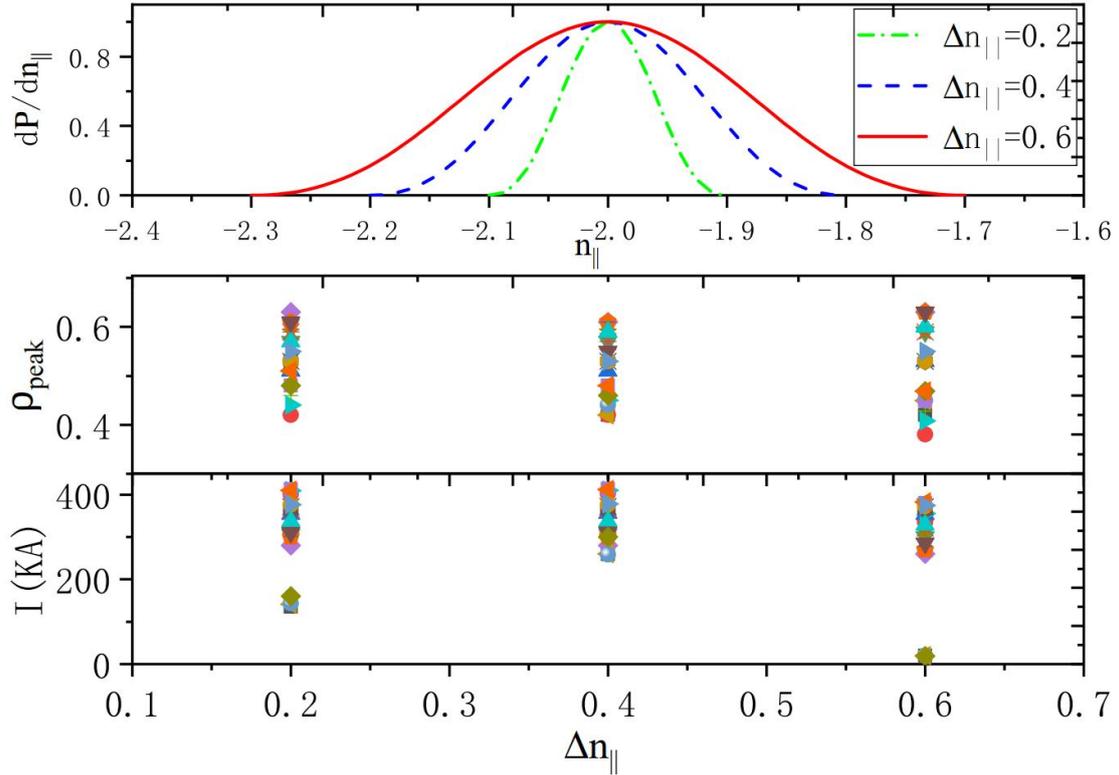

Fig 10. Results of simulations for the CFETR hybrid scenario, with (a) the input profile of $n_{||}$, (b) the peak position of driven current, and (c) the current driven by helicons versus $\Delta n_{||}$ by changing the poloidal angle at a frequency of 1.4GHz and $n_{||}$ of 1.6.

## 4. Summary

In this work, we have demonstrated the results of helicon heating and current drive in a hybrid scenario for CFETR. Utilizing the GENRAY/CQL3D package, a large number of simulations (over 5 000) were conducted, with parametric scans in the antenna's poloidal position, launched parallel refractive index, and wave frequency. The analysis demonstrates that helicon waves exhibit excellent accessibility under conditions of a reactor size, and that higher wave frequency and smaller parallel refractive index lead to slightly improved wave absorption. Moreover, an optimal $n_{||}$ value of approximately 1.6 is identified for the hybrid scenario in CFETR, with helicon achieving a maximum drive efficiency of $2.8\times10^{19} A \cdot W^{-1} \cdot m^{-2}$. This means that it can achieve 70% of the optimal drive current for lower hybrid wave in the same configuration. The best launch position is found to be within a poloidal angle range of 25 degrees to 65 degrees. In addition, it is desirable to have a narrow $n_{||}$ spectrum for wave absorption when operating below the threshold value of $\Delta n_\parallel (\approx 0.6)$, beyond which the impact of $\Delta n_\parallel$ on wave absorption becomes insignificant. This investigation offers valuable insights into the potential implementation of HCD in CFETR and future Demo reactor.

**Acknowledgments**

This work is supported by the National Natural Science Foundation of China (Grants No. 11905109, No. 11905080, and No.11947238), the National Key R&D Program of China (Grants No. 2018YFE0303102 and No. 2017YFE0301702), and the Center for Computational Science and Engineering of Southern University of Science and Technology.